\documentclass[intlimits,twoside,a4paper]{article}

\usepackage[cp1251]{inputenc}

\usepackage[eqsecnum]{cmpj3}

\usepackage{bm}

\articletype{A part of the Special collection to the 100th anniversary of the birth of Ihor Yukhnovskii}

\issue{2025}{28}{2}{23605}
\doinumber{10.5488/CMP.28.23605}

\title[On the diffusion of hard sphere fluids in disordered porous media]%
{On the diffusion of hard sphere fluids in disordered porous media: New extended Enskog theory description%
\author[M. F. Holovko, M. Ya. Korvatska]{M. F. Holovko\orcid{0000-0001-8114-5356},
	M. Ya. Korvatska\orcid{0000-0003-3455-0468}}
\address{
	Institute for Condensed Matter Physics of the National Academy of Sciences of Ukraine,
	1 Svientsitskii St., 79011 Lviv, Ukraine
	}}

\date{Received April 10, 2025, in final form May 24, 2025}

\begin{document}

\maketitle

\begin{abstract}
	We proposed a new extended version of Enskog theory for the description of the self-diffusion coefficient of a colloidal hard-sphere fluid adsorbed in a matrix of disordered hard-sphere obstacles. In a considered approach instead of contact values of the fluid-fluid and fluid-matrix pair distribution functions, we introduced by	input the new functions that include the dependence on the fraction of the volume free from matrix particles and from fluid particles trapped by matrix particles. It is shown that the introduction of	this free volume fraction by the Fermi-like distribution leads to the best agreement between theoretical predictions and computer simulation results [Chang R., Jagannathan K., Yethiraj A.,	Phys. Rev. E, 2004, \textbf{69}, 051101].
	
\end{abstract}

\section{Introduction}

This paper is devoted to the 100th anniversary of the birth of the well-known Ukrainian scientist Igor Yukhnovskii, who was our teacher and founder of the Institute for Condensed Matter Physics of the National Academy of Sciences of Ukraine. One of us (Myroslav Holovko) had the great pleasure and benefit of working with him for nearly 60 years, starting from 1965. 

For the last time, starting from the pioneering paper of Madden and Glandt \cite{MadGlandt88}, much theoretical effort has been devoted to the study of fluid adsorption in disordered porous media. According to the model of Madden and Glandt, a disordered porous medium is presented as a matrix of quenched configurations of randomly distributed obstacles. The specificity of such an approach is connected with the double quenched-annealed averages: the annealed average is taken over all fluid configurations with fixed coordinates of matrix obstacles, and after that, an additional quenched average should be taken for the free energy of the fluid at fixed coordinates of matrix particles over all realizations of the matrix.

One of the simplest models for the fluid in porous media is a hard sphere fluid in a hard sphere matrix. For the analytical description of thermodynamical properties of such a model, Holovko and Dong \cite{HolDong09} have proposed to extend the classical scaled particle theory (SPT) \cite{ReissFrisLeb59} to hard-sphere fluid in porous media. After that the SPT approach for hard sphere fluids in disordered porous media was essentially improved and developed \cite{ChenDong10, PatHol11, HolPat12, HolPatDong12, HolPatDong17}. The approach proposed in \cite{HolDong09}, named SPT1, contains a subtle inconsistency that was eliminated in the approach named SPT2 \cite{PatHol11}. As a result, the first rather accurate analytical expressions were obtained for the chemical potential and pressure of a hard sphere fluid confined in a hard sphere matrix. The obtained expressions include three parameters defining the porosity of the matrix. The first one is the coefficient of geometric porosity
\begin{equation}
	\phi_{0}=1-\eta_{0}
	\label{HolKor1.1}
\end{equation}
characterizing the free volume, which is not occupied by the matrix obstacles, $\eta_{0}=\frac{1}{6}\piup\rho_{0}\sigma_{0}^{3}$ is the part of the volume occupied by the matrix obstacles with the size $\sigma_{0}$ and the density $\rho_{0}={N_{0}}/{V}$. The second parameter 

\begin{equation}
	\phi=\phi_{0}\exp\left\{ -\frac{\eta_{0}\tau}{1-\eta_{0}}\left[ 3(1+\tau)+\frac{9}{2}\frac{\eta_{0}}{1-\eta_{0}}\tau+\frac{1+\eta_{0}+\eta_{0}^{2}}{(1-\eta_{0})^{2}}\tau^{2}\right] \right\}  
	\label{HolKor1.2}
\end{equation}
is the coefficient of probe particle porosity defined by the excess chemical potential of a fluid in the limit of infinite dilution. It characterizes the adsorption of a fluid in an empty matrix. $\tau=\sigma_{1}/\sigma_{0}$, where $\sigma_{1}$ is the size of a fluid hard sphere. The third parameter $\phi^{*}$ is defined by the maximum value of the packing fraction of hard sphere fluid in a porous media. The general expression for $\phi^{*}$ was proposed in \cite{HolPatDong12,HolPatDong17}
\begin{equation}
{\phi^{*}}=\frac{{\phi_{0}}{\phi}}{{\phi_{0}}-\phi}\ln\frac{\phi_{0}}{\phi},
\label{HolKor1.3}
\end{equation}
which is exact for the one-dimensional case and can be considered as an appropriate approximation for higher dimensions.

We note that in the case of simple fluids, the sizes of fluid particles are considerably lower than the sizes of matrix particles, and we can put $\tau=0$. In such a situation $\phi=\phi^*=\phi_{0}$, and the influence of porous media is connected only with the excluded volume of matrix particles described by $\phi_{0}$. However, in the case of colloidal fluid, the sizes of colloidal and matrix particles can be comparable, and in such a situation, we can put $\tau=1$ and $\phi<\phi^{*}<\phi_{0}$. Under such conditions, the effects described by all porosity coefficients are important. In particular, in addition to the effects of excluded volume by matrix particles described by coefficient $\phi_{0}$, the coefficient $\phi$ also includes the effects from fluid particles adsorbed by matrix particles. We would like to emphasize that even in the case of point matrix particles at $\sigma_{0}=0$, $\phi_{0}=1$ and $\phi=\exp(-\eta_{0}\tau^3)\neq1$, where $\eta_{0}\tau^3=\frac{1}{6}\piup\rho_{0}\sigma_{1}^{3}$. Such a situation corresponds to the porous media created by frozen point defects.

The SPT2 is also useful for the description of the static structure of hard sphere fluids in disordered porous media. Especially in \cite{KalHol14} and \cite{HolovKor20}, simple and rather accurate expressions were obtained for the contact values of the fluid-fluid and fluid-matrix pair distribution functions, correspondingly.
\begin{equation}
	g_{11}(\sigma_{11})=\frac{1}{\phi_{0}-\eta_{1}}+\frac32\frac{\eta_{1}+\eta_{0}\tau}{(\phi_{0}-\eta_{1})^{2}}+\frac12\frac{(\eta_{1}+\tau\eta_{0})^{2}}{(\phi_{0}-\eta_{1})^{3}},
	\label{HolKor1.4}
\end{equation}
\begin{equation}
	g_{10}(\sigma_{10})=\frac{1}{\phi_{0}-\eta_{1}}+\frac{3}{1+\tau}\frac{\eta_{1}+\eta_{0}\tau}{(\phi_{0}-\eta_{1})^{2}}+\frac{2}{(1+\tau)^2}\frac{(\eta_{1}+\tau\eta_{0})^{2}}{(\phi_{0}-\eta_{1})^{3}}  ,
	\label{HolKor1.5}
\end{equation}
where $\eta_{1}=\frac{1}{6}\piup\rho_{1}\sigma_{1}^{3}$ is the packing fraction of fluid hard spheres with the size $\sigma_{1}$ and the density $\rho_{1}={N_{1}}/{V}$, $\sigma_{10}=\frac12(\sigma_{0}+\sigma_{1})$. We note that we use the conventional notation \cite{ChenDong10, PatHol11, HolPat12, HolPatDong12, HolPatDong17, KalHol14, HolovKor20}: the index ``1'' denotes the fluid component and the index ``0'' denotes the matrix particles.

The obtained expressions (\ref{HolKor1.4})--(\ref{HolKor1.5}) for the contact values were used by us in \cite{HolovKor20}, similar to the bulk case \cite{ResibLee77}, as a source of input information in the Enskog theory for the description of transport properties of hard sphere fluids in porous media. For this purpose in \cite{HolovKor20}, the hard sphere fluid confined in disordered porous media was considered as a two-component mixture, one of which is quenched and is treated as hard-sphere particles with infinite mass. However, the comparison between the self-diffusion coefficient of hard-sphere fluid in a porous hard-sphere matrix calculated from Enskog theory modified in such a way and computer simulation data obtained in the Yethiraj group \cite{ChangJad04} shows a systematic overestimation of theoretical prediction, which increases with the increasing fraction of matrix particles.

It means that the treatment of quenched matrix particles as the particles with infinite mass is not sufficient for modeling the fluid in porous media. It is very important in the formation of the input information for the extended Enskog theory for the transport properties of fluid in porous media to take into account that matrix particles are immobile and fluid particles are mobile. As a result, the fluid particles can be either hindered in their motion by cages formed by other mobile fluid particles or trapped by the immobile matrix particles. Each of these processes and their interplay are responsible for slowing down the dynamics of the fluid and leading to a complex dynamic behavior of fluid in random porous media~\cite{KuzCosKah01}. In this paper, we show that such complex dynamics of fluid strongly correlates with the ratio $\frac{\phi}{\phi_{0}}$, which characterizes the fraction of the  volume free from matrix particles and also free from fluid particles trapped by matrix particles. In our previous paper \cite{HolovKor20}, we note that such improvement of the description of transport properties of fluid in porous media strongly correlates with the description of thermodynamic properties of hard sphere fluids in random porous media, where the dependence on ${\phi_{0}}$ and ${\phi}$ is also very important.

The paper is organized as follows. In section~\ref{sec:2} we briefly summarize our previous results \cite{HolovKor20} for the diffusion of the hard-sphere fluid in random porous media with a corresponding generalization and improvement. In section~\ref{sec:3}, numerical results are presented together with the comparison with computer simulation data. Finally, we draw conclusions in the last section.

\section{Extended Enskog theory for hard sphere fluids in disordered porous media}
\label{sec:2}

The original Enskog equation for hard sphere fluids was formulated over a hundred years ago~\cite{EnsKung22} as the generalization of the Boltzmann kinetic equation to high densities of hard sphere fluids by including the pressure non-ideal term  as a multiplier. Due to the virial theorem	\cite{YukhHol80}, this multiplier can be presented in the form
\begin{equation}
\frac{P_{1}}{\rho_{1}kT}-1=4\eta_{1}g_{11}(\sigma_{1}),
	\label{HolKor2.1}
\end{equation}
where $P_{1}$ is the pressure of hard sphere fluid, $k$ is the Boltzmann constant, and $T$ is the absolute temperature. $g_{11}(\sigma_{1})$ is the contact value of the pair distribution function, which is introduced in Enskog theory in a semi-empirical way as an input parameter. 

In the Enskog theory, only binary collisions between hard spheres are taken into account, and the collision is considered completely independent and instantaneous	\cite{ResibLee77}. Since each collision process in hard sphere fluid is instantaneous, the self-diffusion coefficient $D_{1}$ is related to the corresponding friction coefficient $\xi_{1}$ via the Einstein relation as 
\begin{equation}
	D_{1}= kT/\xi_{1},  
	\label{HolKor2.2}
\end{equation}
where $\xi_{1}$ is the friction coefficient in the stationary limit.

The presented results can be easily generalized to the hard-sphere mixtures, and after that, we can follow \cite{HolovKor20, ChangJad04} and mimic a hard sphere fluid in a hard sphere matrix by a binary mixture where one component is infinitely massive. As a result, we will have
\begin{equation}
	\xi_{1}=32\left(\frac{kT m_{1}}{2\piup\sigma_{11}^{2}}\right)^{1/2}\left[ \frac{1}{\sqrt{2}}\eta_{1}g_{11}(\sigma_{11})+\frac14\tau(\tau+1)^{2}\eta_{0}g_{10}(\sigma_{10})\right] ,
	\label{HolKor2.3}
\end{equation}
where the expressions for the contact values $g_{11}(\sigma_{11})$ and $g_{10}(\sigma_{10})$ are given by the expressions (\ref{HolKor1.4}) and~(\ref{HolKor1.5}).

In accordance with (\ref{HolKor2.2}), we have the following expression for the self-diffusion coefficient of a hard sphere fluid in a disordered porous medium.
\begin{equation}
	D_{1}/D_{1}^{0}=\frac{\sqrt{2\piup}}{32} \left[ \frac{1}{\sqrt{2}}\eta_{1}g_{11}(\sigma_{11})+\frac14\tau(\tau+1)^{2}\eta_{0}g_{10}(\sigma_{10})\right]^{-1} ,
	\label{HolKor2.4}
\end{equation}
where $D_{1}^{0}= \left( {kT\sigma_{11}^{2}}/{m_{1}}\right) ^{1/2}$,  $m_{1}$ is the mass of the fluid particle. 
For pure hard-sphere fluid $\eta_{0}=0$, and~(\ref{HolKor2.4}) reduces for well-known results for pure hard-sphere fluid \cite{ResibLee77}
\begin{equation}
	D_{1}=\frac{1}{16}\frac{\left( {kT\sigma_{11}^{2}}/{m_{1}}\right) ^{1/2}}{\eta_{1}g_{11}(\sigma_{11})}.
	\label{HolKor2.5}
\end{equation}
However, in the presence of porous media, there is no simple expression like (\ref{HolKor2.1}) for the pressure and in expression (\ref{HolKor2.4}) for the self-diffusion coefficient, the contact value $g_{11}(\sigma_{11}) $ and $g_{10}(\sigma_{10})$ should be changed into a new property $G_{11}(\sigma_{11})$ and $G_{10}(\sigma_{10})$, which should include the dependence on the probe particle porosity $\phi$ and the effect of trapping of mobile fluid particles by immobile matrix particles. In our previous paper	\cite{HolovKor20}, we discussed such a modification only for $g_{10}(\sigma_{10})$. In this paper we also propose such a modification for $g_{11}(\sigma_{11})$. First of all, new functions $G_{10}(\sigma_{10}) $ and $G_{11}(\sigma_{11})$ should include the following ratio  as the multiplier 
\begin{equation}
\frac{\phi_{0}}{\phi}=\exp\left[ \frac{3\tau(1+\tau)\eta_{0}}{1-\eta_{0}}+\frac{9}{2}\frac{\eta_{0}^2}{(1-\eta_{0})^{2}}\tau^2+\frac{1+\eta_{0}+\eta_{0}^{2}}{(1-\eta_{0})^{3}}\eta_{0}\tau^{3}\right] . 
	\label{HolKor2.6}
\end{equation}
The inverse value of this ratio $\frac{\phi}{\phi_{0}}$ characterizes the fraction of volume free from matrix particles and free from fluid particles trapped by matrix particles. Similar to the description of thermodynamic properties~\cite{HolPat12, HolPatDong12, HolPatDong17}, we can put
\begin{equation}
	G_{10}(\sigma_{10})=\frac{\phi_{0}}{\phi}\left[ 
	\frac{1}{1-\eta_{1}/\phi}+
	\frac{3}{1+\tau}\frac{\eta_{1}/\phi_{0}}{(1-\eta_{1}/\phi_{0})^{2}}+ \frac{2}{(1+\tau)^2}\frac{(\eta_{1}/\phi_{0})^{2}}{(1-\eta_{1}/\phi_{0})^{3}}\right] .
	\label{HolKor2.7}
\end{equation}
The first term in (\ref{HolKor2.7}) has a divergence at $\eta_{1}=\phi$, and similar to the thermodynamic consideration in~\cite{HolovKor20}, we change ${1}/{(1-\eta_{1}/\phi)}$ to
\begin{equation}
	\frac{1}{1-\eta_{1}/\phi}=\frac{1}{1-\eta_{1}/\phi_{0}}+ \frac{\eta_{1}(\phi_{0}-\phi)}{\phi_{0}\phi(1-\eta_{1}/\phi_{0})^{2}}+\cdots
	\label{HolKor2.8}
\end{equation}
and $g_{10}(\sigma_{10})$ we change into
\begin{align}
	G_{10}(\sigma_{10}) &= \frac{\phi_{0}}{\phi}\left[ 
	\frac{1}{1-\eta_{1}/\phi_{0}}+ \frac{\eta_{1}(\phi_{0}-\phi)}{\phi_{0}\phi(1-\eta_{1}/\phi_{0})^{2}}+
	\frac{3}{1+\tau}\frac{\eta_{1}/\phi_{0}}{(1-\eta_{1}/\phi_{0})^{2}}  \right. \nonumber\\
	&\left.+ \frac{2}{(1+\tau)^2}\frac{(\eta_{1}/\phi_{0})^{2}}{(1-\eta_{1}/\phi_{0})^{3}}\right] .
	\label{HolKor2.9}
\end{align}
Similar manipulation leads to the change $g_{11}(\sigma_{11})$ into
\begin{equation}
	G_{11}(\sigma_{11})=\frac{\phi_{0}}{\phi}\left[ 
	\frac{1}{1-\eta_{1}/\phi_{0}}+ \frac{\eta_{1}(\phi_{0}-\phi)}{\phi_{0}\phi(1-\eta_{1}/\phi_{0})^{2}}
	+\frac{3}{2}\frac{\eta_{1}/\phi_{0}}{(1-\eta_{1}/\phi_{0})^{2}}+ \frac{1}{2}\frac{(\eta_{1}/\phi_{0})^{2}}{(1-\eta_{1}/\phi_{0})^{3}}\right] .
	\label{HolKor2.10}
\end{equation}
As a result, for the self-diffusion coefficient of a hard-sphere fluid in a hard-sphere matrix, we have a new expression 
\begin{equation}
	D_{1}/D_{1}^{0}=\frac{\sqrt{2\piup}}{32} \left[ \frac{1}{\sqrt{2}}\eta_{1}G_{11}(\sigma_{11})+\frac14\tau(\tau+1)^{2}\eta_{0}G_{10}(\sigma_{10})\right]^{-1} ,
	\label{HolKor2.11}
\end{equation}
which differs from (\ref{HolKor2.4}) only by changes $g_{11}(\sigma_{11}) $ and $g_{10}(\sigma_{10}) $ into $G_{11}(\sigma_{11}) $ and $G_{10}(\sigma_{10}) $ given by the expressions (\ref{HolKor2.10}) and (\ref{HolKor2.9}), correspondingly. 

We should note that the new functions $G_{11}(\sigma_{11})$ and $G_{10}(\sigma_{10})$ are not contact values of pair distribution functions in contrast to $g_{11}(\sigma_{11})$ and $g_{10}(\sigma_{10})$.
\begin{figure}[h]
	\centerline{
		\includegraphics [height=0.42\textwidth]{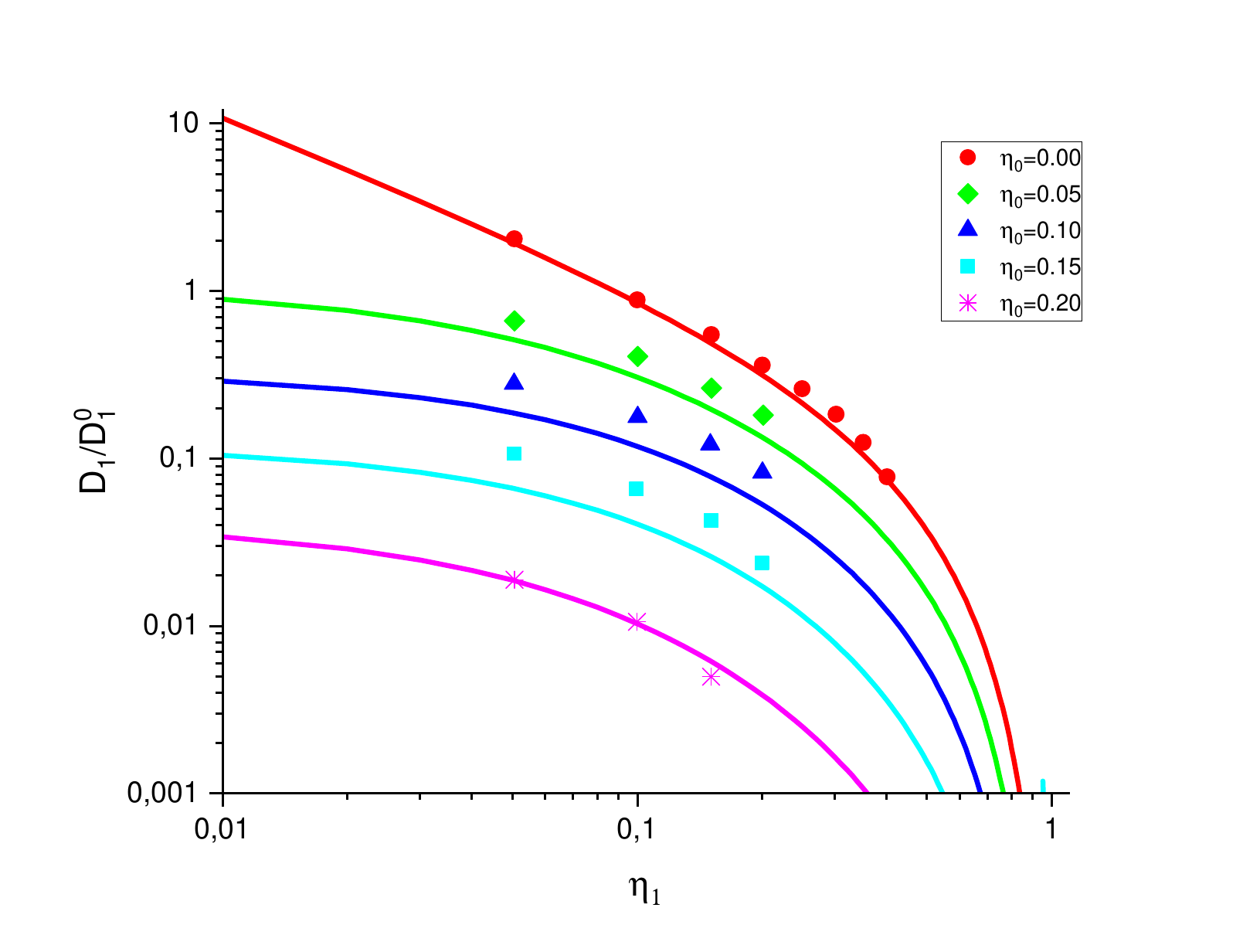}
		\includegraphics [height=0.42\textwidth]{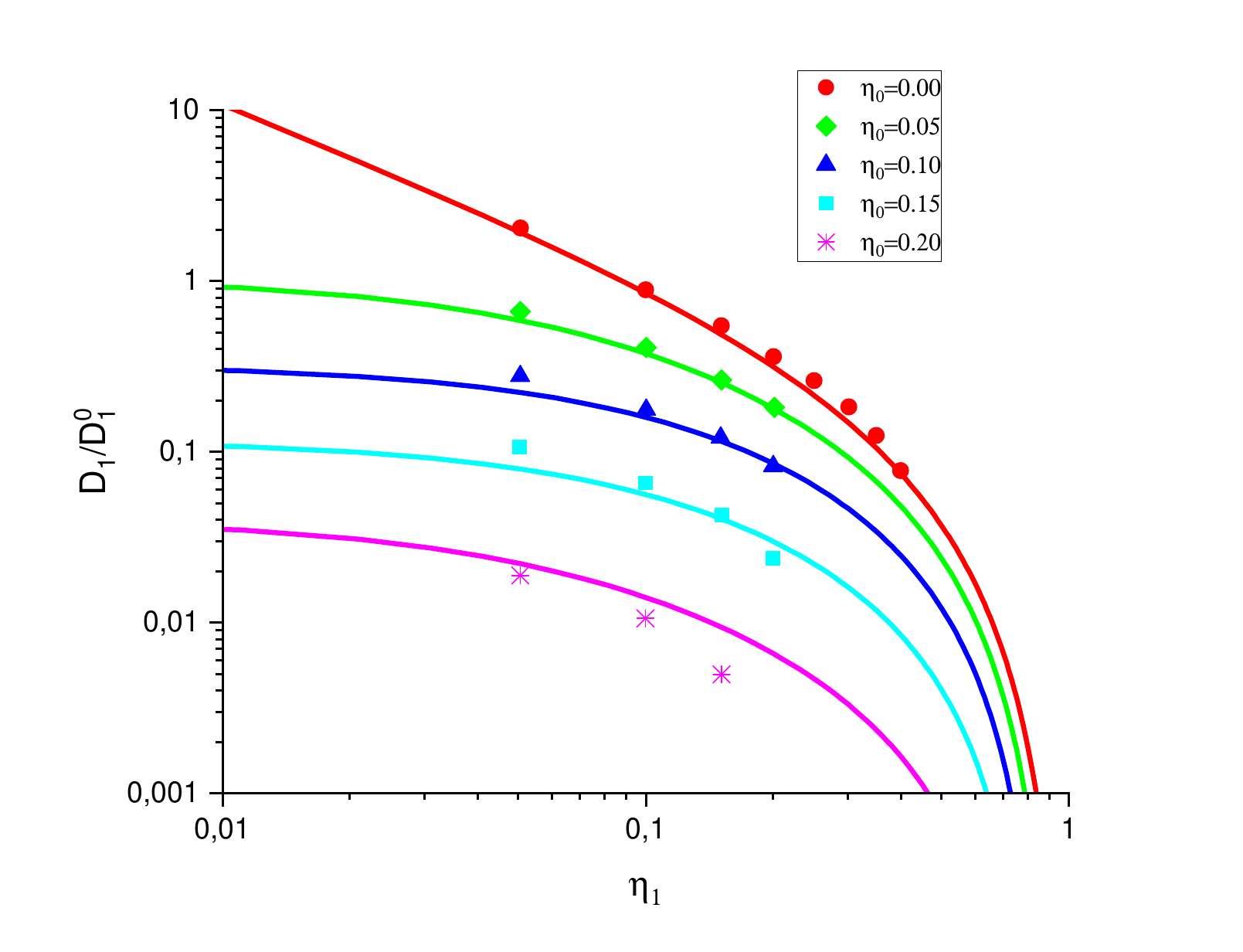}
	}
	\caption{(Colour online) Comparison of the EET prediction equation (\ref{HolKor2.4}) and computer simulation results	\cite{ChangJad04} for the diffusion coefficient $D_{1}$ for hard sphere fluid in hard sphere matrix as a function of fluid packing fraction $\eta_{1}$ for different matrix packing fractions $\eta_{0}$ and for $\tau=1$. The solid line corresponds to the theoretical prediction EET.
		Panel a: Functions $G_{10}(\sigma_{10}) $ and $G_{11}(\sigma_{11}) $ are given by the expressions (\ref{HolKor2.9}) and (\ref{HolKor2.10}), correspondingly. 
		Panel b: Functions $G_{11}(\sigma_{11}) =g_{11}(\sigma_{11})$ and is given by the expressions  (\ref{HolKor1.4}). $G_{10}(\sigma_{10})$ is given by the expressions (\ref{HolKor2.9}).}
	\label{FIG1}
\end{figure}

\section{Results and discussions}
\label{sec:3}

In this section we  discuss the effect of the packing fraction of hard sphere fluid $\eta_{1}$ and the packing fraction of hard-sphere porous media $\eta_{0}$ on the self-diffusion coefficient $D_{1}$. To this end, we use the expression (\ref{HolKor2.11}) obtained in the previous section in the framework of the extended Enskog theory~(EET). The generalized functions $G_{10}(\sigma_{10})$ and $G_{11}(\sigma_{11})$ are given by the expressions (\ref{HolKor2.9}) and~(\ref{HolKor2.10}), correspondingly. Figure~\ref{FIG1}a depicts the self-diffusion coefficient $D_{1}$ as a function of $\eta_{1}$ at a fixed value of $\eta_{0}$ predicted from generalized Enskog theory for the case $\tau=1$ at different values of $\eta_{0}$. The results of computer simulations obtained in the Yethiraj group	\cite{ChangJad04} are also presented for comparison. As we can see,  with an increase of $\eta_{1}$ or $\eta_{0}$, the self-diffusion coefficient $D_{1}$ decreases monotonously. We have a more or less correct agreement between theoretical prediction and computer simulation data only for the bulk case ($\eta_{0}=0$) and for a sufficiently dense matrix packing fraction $\eta_{0}=0.2$. However, for intermediate values $\eta_{0}=0.05$, 0.1, and 0.15, the EET at all $\eta_{1}$ slightly reestimates the value of $D_{1}$. For $\eta_{0}=0$, the result is not strange because for the bulk case ($\eta_{0}=0$) the EET reduces to a standard Enskog theory, which is correct for the bulk case	\cite{ResibLee77}. The correct agreement between the theoretical result and computer simulation data for the dense matrix at $\eta_{0}=0.2$ suggests a possible way for improvement of the EET for intermediate values of $\eta_{0}$, which will be considered later.
 For comparison, in figure~\ref{FIG1}b we repeat the result from our previous paper~\cite{HolovKor20}, in which $G_{11}(\sigma_{11})=g_{11}(\sigma_{11})$ is defined by the expression (\ref{HolKor1.4}) and $G_{10}(\sigma_{10})$ is taken in the form (\ref{HolKor2.9}). We note that in this paper, in contrast to our previous paper \cite{HolovKor20}, we continue the calculated curves to lower values of $\eta_{1}$ up to $\eta_{1}=0.01$. As we can see at small $\eta_{1}$, the  results in both figures~\ref{FIG1}a and~\ref{FIG1}b coincide more or less, and theory overestimates the values of $D_1$ at small $\eta_{1}$. It means that at small fluid parameter $\eta_{1}$, the function $G_{10}(\sigma_{10})$ dominates and slightly underestimates the values of the diffusion coefficient. For higher density of fluid, the function $G_{11}(\sigma_{11})$ underestimates the values of the diffusion coefficient as well, and the calculated values of the diffusion coefficient $D_{1}$ are underestimated at all values $\eta_{1}$. However, the change from $G_{11}(\sigma_{11})$ to $g_{11}(\sigma_{11})$ underestimates the role of fluid-fluid correlation for the diffusion coefficient. As a result, due to compensation inaccuracy of matrix-fluid and fluid-fluid correlation in terms of $G_{10}(\sigma_{10}) $ and $g_{10}(\sigma_{10}) $, the change of $G_{11}(\sigma_{11}) $ to $g_{11}(\sigma_{11})$ with an increase of the fluid density parameter $\eta_{1}$ as we can see from figure~\ref{FIG1}b leads to an improvement of the theoretical description of diffusion coefficient $D_{1}$. 
 
\begin{figure}[h]
 	\centerline{
		\includegraphics [height=0.42\textwidth]{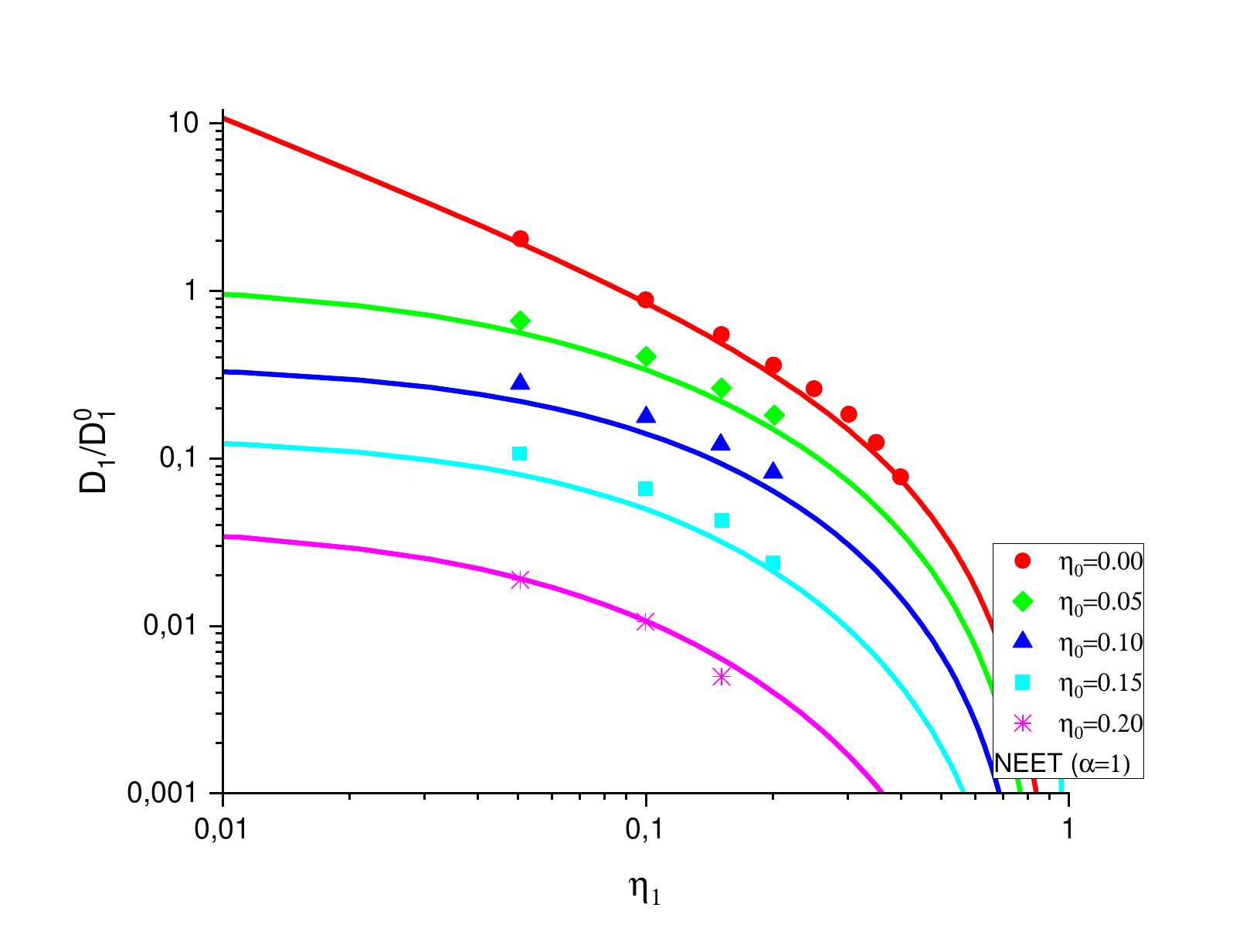}
 		\includegraphics [height=0.42\textwidth]{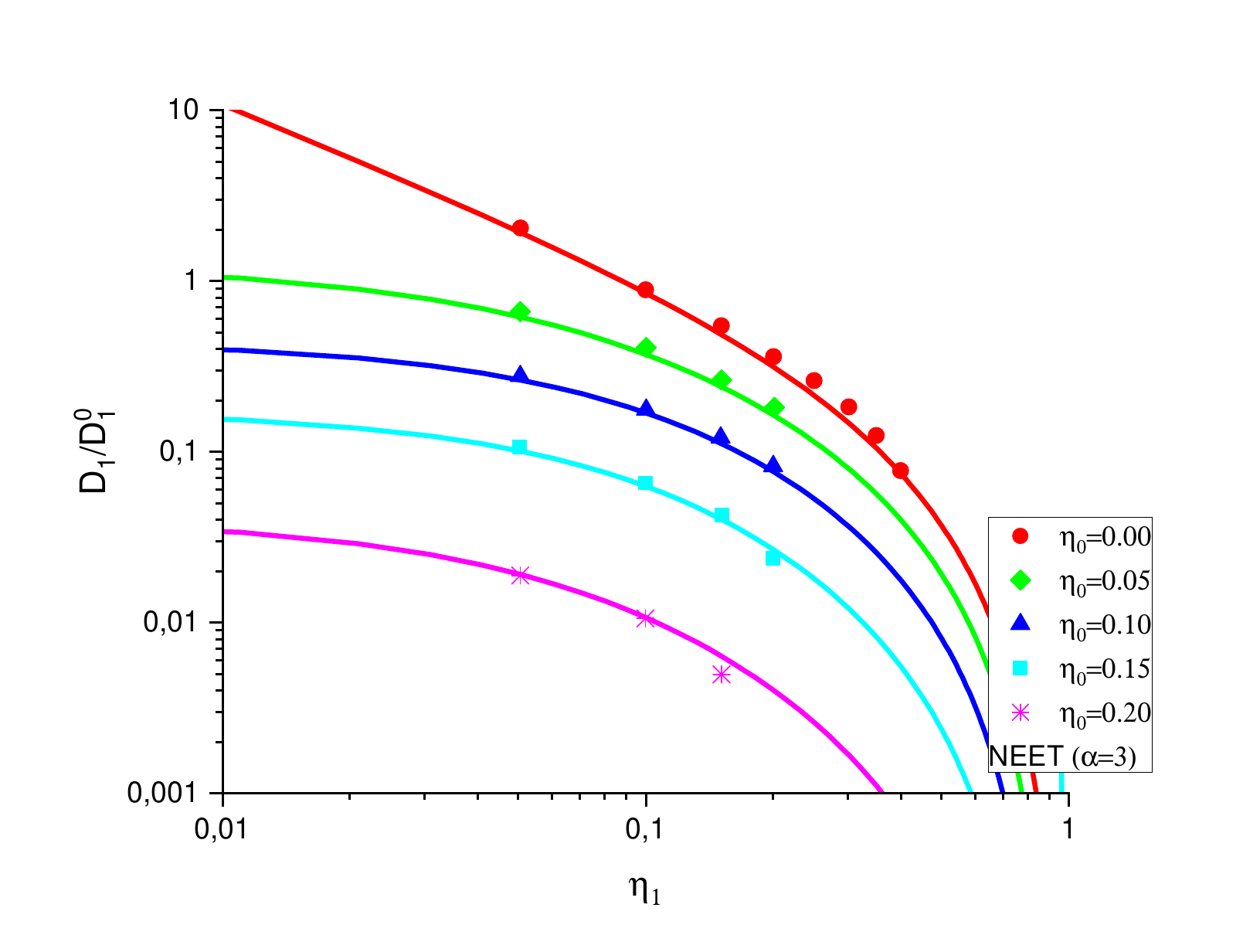}
 	}
 	\caption{(Colour online) Comparison of the NEET  prediction equation (\ref{HolKor2.4}) and computer simulation results	\cite{ChangJad04} for the diffusion coefficient $D_{1}$ for hard sphere fluid in hard sphere matrix as a function of fluid packing fraction $\eta_{1}$ for different matrix packing fractions $\eta_{0}$ and for $\tau=1$. Functions $G_{10}(\sigma_{10})$ and $G_{11}(\sigma_{11})$ are given by the expressions (\ref{HolKor3.2}) and (\ref{HolKor3.3}), correspondingly. The solid line corresponds to the theoretical prediction NEET.
 	Panel a: $\alpha=1$.
 	Panel b: $\alpha=3$.}
 	\label{FIG2}
 \end{figure}

 Now, we return to the discussion of the problem of the EET for the intermediate values of $\eta_{0}$, which appears from the consideration of figure~\ref{FIG1}a. From the analyses of figure~\ref{FIG1}a, we see that some kind of restriction should be taken into account for the parameter $\frac{\phi}{\phi_{0}}$, which was introduced by us for the function $G_{10}(\sigma_{10})$ and $G_{11}(\sigma_{11})$ defined by the expressions (\ref{HolKor2.9}) and (\ref{HolKor2.10}). In the present paper, this multiplier is modified by the function of $\eta_{0}$ and $\frac{\phi}{\phi_{0}}$, which change from 1 near the bulk case (at small $\eta_{0}$) to $\frac{\phi}{\phi_{0}}$ near the saturated value $\eta_{0}^*$, which is defined by the porous material and specific fluid. For the considered case, as we can see from figure~\ref{FIG1}a, to reach an agreement with computer simulation data, it is possible to fix $\eta_{0}^*=0.4$. In this paper we suggest taking the discussed multiplier in the Fermi-like form. It means that, according to this correction, the inverse value of the volume free from the matrix particles and from fluid particles trapped by the matrix particles will have Fermi-like for dependence on $\eta_{0}$ instead of $\frac{\phi_{0}}{\phi}$ considered in the previous section
 \begin{equation}
 	\frac{\phi_{0}}{\phi}\to \frac{1}{1+\big(\frac{\phi}{\phi_{0}}-1\big)\exp{[\alpha(\eta_{0}-\eta_{0}^*)]}},
 	\label{HolKor3.1}
 \end{equation}
where we have introduced the parameter $\alpha$. We consider this parameter as the fitting parameter.  It makes it possible to control the role of the packing fraction of matrix particles $\eta_{0}$ in order to improve the agreement between theoretical description and computer simulation data. We note that a similar problem also appears  in the theory of the electrical double layer, where, due to saturation of the adsorption ions on the charged surface, there arises a problem of changing the Poisson--Boltzmann distribution into the Poisson--Fermi distribution \cite{BokCapH18}. We note that near the bulk case, the packing fraction of porous material $\eta_{0}$ is small and the multiplier parameter (\ref{HolKor3.1}) is equal to 

\begin{equation}
	 \frac{1}{1+\big(\frac{\phi}{\phi_{0}}-1\big)\exp{(-\alpha}\eta_{0}^*)}\approx1,
	\label{HolKor3.2}
\end{equation}
and we return to the expression (\ref{HolKor2.4}) for the EET for hard spheres in porous media. In the opposite case when parameter $\eta_{0}=\eta_{0}^*$, the multiplier parameter (\ref{HolKor3.1}) is equal to
\begin{equation}
	\frac{1}{1+\big(\frac{\phi}{\phi_{0}}-1\big)}=\frac{\phi_{0}}{\phi},
	\label{HolKor3.3}
\end{equation}
and we return to the case of EET considered in the previous section. Having changed the ratio~$\frac{\phi_{0}}{\phi}$ to~(\ref{HolKor3.1}), the functions $G_{11}(\sigma_{11})$ and $G_{10}(\sigma_{10}) $ introduced by us in the EET approach in the form of (\ref{HolKor2.9}) and~(\ref{HolKor2.10}) will be given by the expressions:
 \begin{eqnarray}
 	\nonumber
 	G_{10}(\sigma_{10})&=&\frac{1}{1+\big(\frac{\phi}{\phi_{0}}-1\big)\exp{[\alpha(\eta_{0}-\eta_{0}^*)]}}\left[ 
 	\frac{1}{1-\eta_{1}/\phi_{0}}\right.\nonumber\\
 	&+&\left. \frac{\eta_{1}(\phi_{0}-\phi)}{\phi_{0}\phi(1-\eta_{1}/\phi_{0})^{2}}+
 	\frac{3}{1+\tau}\frac{\eta_{1}/\phi_{0}}{(1-\eta_{1}/\phi_{0})^{2}}+ \frac{2}{(1+\tau)^2}\frac{(\eta_{1}/\phi_{0})^{2}}{(1-\eta_{1}/\phi_{0})^{3}}\right] .
 	\label{HolKor3.4}
 \end{eqnarray}
 Similar manipulation leads to the change $g_{11}(\sigma_{11})$ into:
 \begin{eqnarray}
 	\nonumber
 	G_{11}(\sigma_{11})&=&\frac{1}{1+\big(\frac{\phi}{\phi_{0}}-1\big)\exp{[\alpha(\eta_{0}-\eta_{0}^*)]}}\left[ 
 	\frac{1}{1-\eta_{1}/\phi_{0}}\right.\nonumber\\
 	&+&\left. \frac{\eta_{1}(\phi_{0}-\phi)}{\phi_{0}\phi(1-\eta_{1}/\phi_{0})^{2}}
 	+\frac{3}{2}\frac{\eta_{1}/\phi_{0}}{(1-\eta_{1}/\phi_{0})^{2}}+ \frac{1}{2}\frac{(\eta_{1}/\phi_{0})^{2}}{(1-\eta_{1}/\phi_{0})^{3}}\right] .
 	\label{HolKor3.5}
 \end{eqnarray}
 This version of the extension of Enskog theory for hard-sphere fluid in porous media we consider as a new extended Enskog theory (NEET). For the usual Fermi-like distribution, we put $\alpha=1$. The comparison of the NEET prediction and computer simulation results \cite{ChangJad04} for the self-diffusion coefficient~$D_{1}$ for the hard-sphere fluid in the hard-sphere matrix as a function of fluid packing fraction $\eta_{1}$ for the different matrix fractions $\eta_{0}$ and for $\tau=1$ is presented in figure~\ref{FIG2}a. As we can see, the agreement between theoretical and simulation results is improved in the framework of NEET. We have a good agreement between theoretical and computer simulation results for $\eta_{0}=0$ and $\eta_{0}=0.2$. However, for other values of $\eta_{0}$, the theory slightly underestimates the computer simulation data. In order to improve a theoretical description, we can use parameter $\alpha$. The corresponding results for $\alpha=3$ are presented in figure~\ref{FIG2}b.

 \newpage
\section{Conclusions}

In this paper we propose a new extended version of the Enskog theory for hard sphere fluids in disordered porous media. In the developed approach, we introduce the new functions $G_{10}(\sigma_{10})$ and $G_{11}(\sigma_{11})$ as generalizations of fluid-matrix and fluid-fluid contact values, which are used as an input information in the EET. The functions $G_{10}(\sigma_{10})$ and $G_{11}(\sigma_{11})$ include the dependence on the fraction of the volume free from matrix particles and fluid particles trapped by matrix particles and are defined by the ratio of geometrical porosity $\phi_{0}$ to thermodynamical porosity $\phi$. The proposed extended version of Enskog theory is used for the description of the self-diffusion coefficient of hard-sphere fluid in disordered porous media. It is shown that the introduction of the dependence of the multiplier on the ratio $\frac{\phi_{0}}{\phi}$ in Fermi-like form (\ref{HolKor3.1}) into the dependence of the function $G_{10}(\sigma_{10})$ and $G_{11}(\sigma_{11})$ on the matrix packing fraction $\eta_{0}$ leads to the best agreement between theoretical prediction results and computer simulation data obtained in group Yethiraj	\cite{ChangJad04} for all considered values of fluid particle and matrix particle packing fractions $\eta_{1}$ and $\eta_{0}$, correspondingly. The Fermi-like form (\ref{HolKor3.1}) has two additional parameters, $\alpha$ and $\eta_{0}^*$, which can be considered as adjustables. The parameter $\alpha$ regulates the influence of the packing fraction of matrix particles $\eta_{0}$ on the transport properties of hard-sphere fluid in porous media. The second parameter, $\eta_{0}^*$, corresponds to the saturated value of the packing fraction $\eta_{0}$ of porous media near which due to the percolation process of the fluid in porous media, the transport properties sharply change \cite{BunHav91}. The formed fluid clusters are fractal, the diffusion will be anomalous, and the description should be modified.

In this paper we focus on the description of the self-diffusion coefficient of hard-sphere fluid in porous media. For the description of more real systems, an additional interaction should be taken into account. To this end, we plan to improve the theoretical treatment in order to combine the Langevin equation with the application of the Mori projector operator method \cite{Boon80} and mode-coupling theory approach \cite{GotSjo92}.

We also intend to generalize the results obtained in this paper to patchy colloids in disordered porous media. The previous results in approximations considered in \cite{HolovKor20} were already presented in \cite{HolovKor21}.

\ukrainianpart

\title{Про дифузію твердокулькового плину в невпорядкованих пористих середовищах: нова розширена теорія Енскога}%
\author{М. Головко, М. Корвацька}
	\address{
		Інститут фізики конденсованих систем Національної академії наук України, вул. Свєнціцького, 1, 79011 Львів, Україна
}

\makeukrtitle
	
	\begin{abstract}
		Ми запропонували нову розширену версію теорії Енскога для опису коефіцієнта самодифузії плину твердих сфер в невпорядкованих пористих середовищах. У розглянутому підході замість контактних значень парних функцій розподілу ``плин-плин'' та ``плин-матриця'' ми ввели нові функції, які включають залежність від частки об’єму, вільного від частинок матриці та від частинок плину, захоплених частинками матриці. Показано, що введення цієї вільної об’ємної частки за допомогою Фермі-подібного розподілу призводить до найкращої узгодженості між теоретичними прогнозами та результатами комп’ютерного моделювання [Chang R., Jagannathan K., Yethiraj A., Phys. Rev. E, 2004, \textbf{69}, 051101].
		\keywords  
		 твердокульковий флюїд, невпорядковані пористі середовища, теорія масштабної частинки, розширена теорія Енскога, коефіцієнт самодифузії
		\end{abstract}
  \end{document}